\documentclass[sigconf]{acmart}

\AtBeginDocument{%
	\providecommand\BibTeX{{%
			\normalfont B\kern-0.5em{\scshape i\kern-0.25em b}\kern-0.8em\TeX}}}

\setcopyright{acmcopyright}
\copyrightyear{2018}
\acmYear{2018}
\acmDOI{10.1145/1122445.1122456}

\acmConference[Woodstock '18]{Woodstock '18: ACM Symposium on Neural
	Gaze Detection}{June 03--05, 2018}{Woodstock, NY}
\acmBooktitle{Woodstock '18: ACM Symposium on Neural Gaze Detection,
	June 03--05, 2018, Woodstock, NY}
\acmPrice{15.00}
\acmISBN{978-1-4503-XXXX-X/18/06}

\begin{document}
	
	\title{States of confusion: Eye and Head tracking reveal surgeons' confusion during arthroscopic surgery}
	
	\author{Benedikt Hosp}

	\authornotemark[1]
	\affiliation{%
		\institution{Human-Computer Interaction, University of Tübingen}
		\streetaddress{Sand 14}
		\city{Tübingen}
		\country{Germany}
		\postcode{72076}
	}
	\email{benedikt.hosp@uni-tuebingen.de}
	
	\author{Myat Su Yin}
	\affiliation{%
		\institution{Mahidol-Bremen Medical Informatics Research Unit, Faculty of Information and Communication Techology}
		\city{Nakhon Pathom}
		\country{Thailand}}
	\email{myat.su@mahidol.ac.th}

	\author{Peter Haddawy}
	\affiliation{%
		\institution{Faculty of ICT, Mahidol University}
		\city{Nakhon Pathom}
		\country{Thailand}\\
		\institution{University of Bremen}
		\city{Bremen}
		\country{Germany}}
	\email{peter.had@mahidol.ac.th}
	
	\author{Ratthapoom Watcharopas}
	\affiliation{%
		\institution{Faculty of medicine, Ramathibodi Hospital, Mahidol University}
		\city{Bangkok}
		\country{Thailand}}
	\email{poom911@hotmail.com}
	
	\author{Paphon Sa-ngasoongsong}
	\affiliation{%
		\institution{Faculty of medicine, Ramathibodi Hospital, Mahidol University}
		\city{Bangkok}
		\country{Thailand}}
	\email{paphonortho@gmail.com}
	
	\author{Enkelejda Kasneci}
	\affiliation{%
		\institution{Human-Computer Interaction, University of Tübingen}
		\streetaddress{Sand 14}
		\city{Tübingen}
		\country{Germany}}
	\email{enkelejda.kasneci@uni-tuebingen.de}

	\renewcommand{\shortauthors}{Hosp et al.}
	
	\begin{abstract}
	During arthroscopic surgeries, surgeons are faced with challenges like cognitive re-projection of the 2D screen output into the 3D operating site or navigation through highly similar tissue. Training of these cognitive processes takes much time and effort for young surgeons, but is necessary and crucial for their education. In this study we want to show how to recognize states of confusion of young surgeons during an arthroscopic surgery, by looking at their eye and head movements and feeding them to a machine learning model. With an accuracy of over 94\% and detection speed of 0.039 seconds, our model is a step towards online diagnostic and training systems for the perceptual-cognitive processes of surgeons during arthroscopic surgeries.
	\end{abstract}

\ccsdesc[500]{Human-centered computing~User models}
\ccsdesc[500]{Applied computing~Health informatics}
\ccsdesc[300]{Human-centered computing~Ubiquitous and mobile computing systems and tools}
	
	\ccsdesc[500]{Computer systems organization~Embedded systems}
	\ccsdesc[300]{Computer systems organization~Redundancy}
	\ccsdesc{Computer systems organization~Robotics}
	\ccsdesc[100]{Networks~Network reliability}
	
	\keywords{gaze, head, eye, tracking, machine learning, random forest, surgery, medicine, confusion}
	
	\begin{teaserfigure}
		\includegraphics[width=\textwidth]{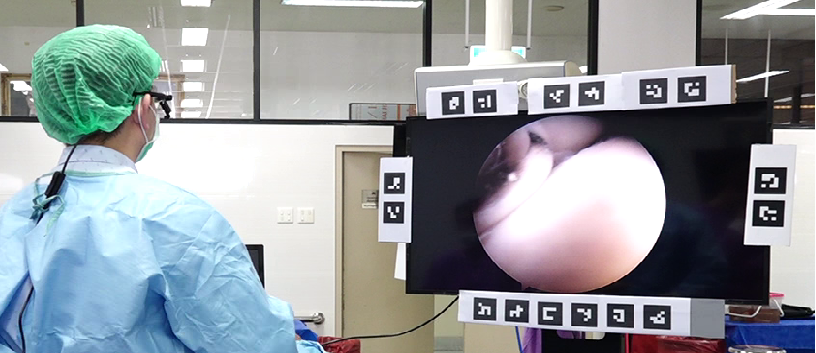}
		\caption{Operating site with surgeon in front of 4k-screen, showing output of arthroscope.}
		\label{fig:teaser}
	\end{teaserfigure}
	
	\maketitle
	
	\section{Introduction}


	Advancements in computer science have typically been a motor for new applications in fields like medicine. Next to classical imagery techniques like magnetic resonance imaging (MRI) \cite{hohne20123d} or arthroscopy \cite{ike2021arthroscopy}, nowadays, the interaction between surgeons and their patients or instruments are increasingly being investigated. There are a lot of new sources of information, e.g. about the vital parameters of the patient or new perspectives/views of the operating site, which are shown to the physicist. They are all meant to improve the work of the surgeon. However, all these new advancements come with a certain level of complexity. Surgeons need to learn how to operate and benefit from this applications. For example, in arthroscopy, the surgeon needs to transfer the 2D image on the scope output into the 3D tissue of the patient. Information is shown on the screen, but navigation takes place on the operating site with a multidimensional instrument. This translation already poses a challenge. 
	
	Even in medical image reading, Brady et al. \cite{brady2017error} estimated that the miss rate for interpreting the results correctly, may be up to 30\% in some ares of radiology. For arthroscopy there is no such study, but arthroscopic surgery is a much more complex procedure than image reading, as surgeons are usually under time and success pressure, while working with patients and the stimulus is constantly and dynamically changing.
	Therefore, ways to teach surgeons to use these new technologies optimally, are as important as the developments of such. This is where human computer interaction comes into play. 
	Methods of human-computer interaction find their way into the world of medicine. Indeed, there are multiple goals to pursue. Besides, i.e. touchless interaction techniques \cite{mewes2017touchless}, the recognition of strategies of surgeons during an operation \cite{sodergren2010hidden} are investigated. 
	The recognition of skill \cite{speidel2006tracking,ahmidi2010surgical,wu2021cross,yin2020study} or states of confusion \cite{stillman2000bedside,zhou2018confusion} of surgeons play a central role in interaction design, as they can help to draw a picture of a surgeons' skills and to find weak-spots which need to be focused on in training. This is done to maximize the output of surgeons and to improve their training. 
	Along with confusion, often frustration or disengagement are involved, if the confusion lasts for too long \cite{d2014confusion}. Pachman et al. \cite{pachman2016eye} summarized different approaches of the last few years and show that multiple ways of detection have been tried, e.g. facial expressions \cite{zeng2008survey,mcdaniel2007facial} or learners' postures \cite{d2009automatic}. D'Mello et al. \cite{d2009automatic} postulated that models based on a single source had high error rates. Thus, later research focused on multiple sources to detect confusion \cite{} but could not be fully automated, as external judges needed to be involved \cite{d2010multimodal}. Most often, surgeons need both of their hands for the operation. So new information and interaction techniques need to focus on other modalities than the surgeons hands. One way to address this, is the use of eye tracking technology. This technique can either be directly used as interaction method \cite{mewes2017touchless} or as information provider about the skill or current state of the surgeons themselves. 
	And as these devices are getting more ubiquitous, faster and more accurate, there are ever new possibilities to study the gaze behavior of the subject. Eye tracking can serve as a perceptual-cognitive diagnosis system. The interest in using eye tracking as a research method in medicine is growing rapidly (for an overview see Lévêque et al \cite{leveque2018state}).
	
	There are even studies that focus on assessment of the impact of training with eye tracking, too \cite{wilson2011gaze,vine2012cheating,krupinski2013characterizing}. Wilson et al. \cite{wilson2011gaze}, i.e found significant differences in completion time when showing young surgeons a video with the gaze signal of an expert during laparoscopy, compared to only showing the plain video of the surgery or allowing a free viewing phase. There are plenty of such studies, showing that 
	the findings of gaze behavior studies can even be used to optimize and/or shorten the training surgeons need to go through. 
	While eye tracking devices are getting faster and ubiquitous, they produce more data, too. On the one hand more data means more usable information, but on the other hand there is a rise in complexity, too. With more data, there can be more inter-dependencies which are hard to understand and handle, especially with traditional techniques  like AOI intersection counts \cite{mackert2013understanding,almansa2011association,kok2015case,kelly2016development,manning2006radiologists}.
	To allow the analysis of such big data to be much more complex, there is another very important advancement in computer science that has an heavy impact on medicine. Artificial intelligence is applied in a variety of applications in medicine \cite{szolovits2019artificial,holzinger2019causability,hamet2017artificial,ramesh2004artificial}. The ever new potentials of machine learning and especially deep learning enabling even more complex tasks to be solved and more data to be analyzed.
	
	In this work we are focusing on the analysis of data from 15 participants during arthroscopic surgery with so called soft-cadavers. During arthroscopy, the surgeon is mainly focusing on the output of the arthroscope, which shows a plane 2D view of the arthroscopic camera inside the portal hole of the patient. Surgeons need to rely on these images, while they navigate through tissue and bones. A young surgeon with low experience may get confused during navigation, since the structures look pretty similar for untrained surgeons. Expert surgeons can rely on their experience and know which visual clues they can use for navigation. In order to optimize the training of young surgeons, we introduce a real time ready confusion detection model, that recognizes states of confusion of surgeons during arthroscopic surgeries. With the combination of eye tracking, head tracking and machine learning methods, we present a highly accurate and fast classification model. Detections of such model can be used to find weak-spots of surgeons in real time and signal assistive actions to be made.
	

	\section{Method}
	
	\subsection{Data collection}

	We collected data of 15 surgeons who are all either members of the Orthopedic Department in the Faculty of Medicine from Mahidol University, Thailand or in the Orthopedics Surgery Residency Program. All subjects were wearing a TobiiGlasses 2 eye tracker (running at 100 Hz) during an arthroscopic surgery of the shoulder on a soft cadaver. The cadaver was placed in front of the surgeon and four feet further away we placed a 4k-screen which shows the output of the scope. During the navigation from the portal hole to the operating side, surgeons were telling verbally where they are and where they go to. They also told when they are confused. Which means they can either not tell their current position inside the joint or how to continue for sure. In relation to the beginning of the operation, we measured these points of time, where the surgeon told to be unsure/confused.

	\subsection{Feature space}
	
	At first we synchronized the eye tracking data with the timing data, by adjusting their timestamps to start at the same time relatively to the start of the surgery. This allows us to find the points of time of confusion inside the eye tracking data. In a next step we cut out a window around every confusion point (+/- one second before and after the event). These pieces of data are considered as "confusion event" samples and the remaining data with no confusion event as "no event" samples.
	
	Each sample contains the following features:
	
	\begin{itemize}
		\item point of regard (x, y)
		\item pupil position (average of both eyes)
		\item pupil diameter (average of both eyes)
		\item gyroscope (x, y, z)
		\item accelerometer (x, y, z)
	\end{itemize}

	\subsection{Classification}
	
	To build a random forest model, we split the samples into training and test data set. This is done in a participant wise manner, which means, if a subject is picked to belong to training set, all of their samples belong to training set. We need to do this, as the model would otherwise learn person-specific, so called idiosyncratic, features (for further information, see \cite{hosp2020eye}).  We followed two different approaches, for testing with unseen data.

	The first approach follows a 2/3-strategy. We randomly pick 2/3 of the subjects for training and count the number of confusion event samples for each. Afterwards, we collect the same amount of "no event" samples from the same subjects. This means for our training set we have the same amount of confusion event samples as no event samples. This firstly leads to a balanced training set (50\% confusion event samples and 50\% no event samples) and secondly, to a chance-level of 50\%, which allows easy interpretation of the results later.
	
	In the second approach, we want to see whether a cross-validation during the training would optimize the results. Thus, we split the training set data by a 5-fold cross validation, which means in every run 1/5 of the data (of the training set) is picked to validate/optimize the model, while 4/5 of the data are used for training the model. 
	After each run, we use the samples of the remaining 1/3 subjects (n=5) to test the classification performance with unseen data.
	
	As we want to use our model in an online fashion, we need to test the classification accuracy (with unseen data) and the classification speed as well. We show the online computability by creating a queue, which consists of n=2000 samples. In our test we keep reading the gaze signal and add one sample to the queue in each step, while the oldest sample is kicked out of the queue. This means at every state the queue has a total of n=2000 samples. The average of each of the features of all samples inside the queue is now computed. These values are now representing the current content of the queue, which we call delta sample. This delta sample is now given to the trained random forest model and to classify it as "confusion sample" or "no confusion sample". To infer the average performance time, we measure the computation time of 100 single runs and calculate the average performance time.

	\section{Results}
	
	Out of 1,266,758 samples, we have 7103 samples with an confusion event and 1,259,655 samples with no event. Out of these samples we collect 7103 confusion samples and 7103 no confusion samples. In every run, we randomly pick 1,000 samples of both to predict their class. The other samples are used for training.
	
\begin{figure}
		\includegraphics[width=0.5\textwidth]{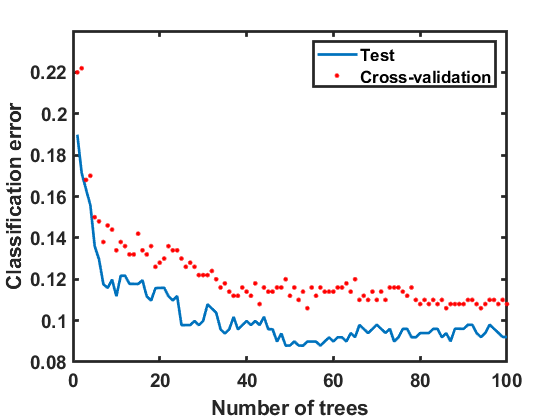}
		\caption{Validation with test data vs. validation with cross-validation as function of number of learners.}
		
		\label{fig:testvscv}
	\end{figure}

	We tested our approach - by randomly assigning training and testing data like aforementioned, in 100 times. 
	The average accuracy of the random forest model is 94.2\%. According to the accuracy, the average misclassification cost / loss is 0.0595. Figure \ref{fig:testvscv} shows the development of the loss over all runs as a function of the number of trained trees. The differences are small but noticeable. The approach with test data set is performing a little bit better than the cross-validation approach. Test set approach reaches the best performance of the cross-validation approach (~0.11) already with about 25-30 trees. The optimal loss value for the test approach is reached at ~50 trees with a misclassification cost of ~0.085.
	
\begin{figure}
		\includegraphics[width=0.4\textwidth]{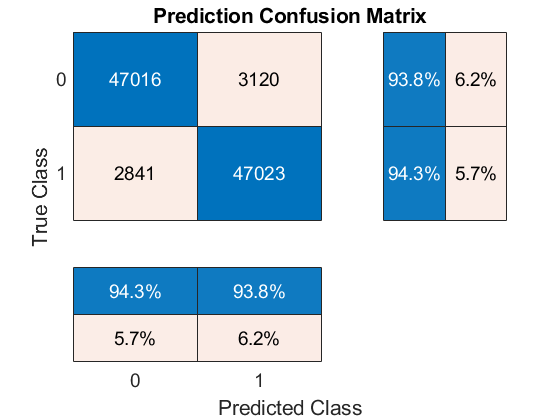}
		\caption{Confusion matrix showing number of correct and falsely predicted samples.}
		
		\label{fig:confMatrix}
	\end{figure}

	Figure \ref{fig:confMatrix} shows the confusion matrix which contains the predictions of all 100 runs. In total we have ~50,000 samples for each class. Of class 0 (no event), 47,016 samples out of 50,136 samples were predicted correctly and 3,120 as confusion event samples. Similarly, for class 1 (confusion event), the model predicted 47,023 samples correctly as confusion event and 2,841 samples wrongly as no event. This result is supported, by the average accuracy over all 100 runs of 94.2\%.

	To measure the performance speed of the model, we measured the computing time of every of the 100 runs. On average the prediction takes 0.039 seconds. This corresponds to a frame rate of ~25 fps.

	\section{Discussion}
	
	In this work we presented a random forest model that is able to classify states of confusions of surgeons during an arthroscopic surgery of the shoulder with an accuracy of over 94.2\%, by taking only 9 features of eye and head movement into account. In our calculations, the model was able to provide a prediction of the content of a queue containing n=2000 samples (2 seconds of samples) in 0.039 seconds. This corresponds to the temporal resolution of common head mounted eye trackers which run at a frame rate between 25-30 fps. The speed may need to be optimized, to allow the application to higher paced field cameras. But in the scenario of a surgery, the speed is not a crucial part, rather, a high detection rate is important. With the detection of confusion states, one can help surgeons to proceed, either pointing out visual clues, which may be used by expert surgeons to navigate, or drawing arrows on the output of the arthroscope which tells the surgeon where to navigate next. Another possible usage of the knowledge of states of confusion can be to augment the whole output by describing the scene by segmenting and labeling each bone or tissue. Or simply name the shown parts in the output. There are multiple ways of supporting the confused surgeon. Depending on the state of expertise, the level of support may be chosen, to allow different skilled surgeons, to train their different weak spots.
	
\begin{figure}
		\includegraphics[width=0.4\textwidth]{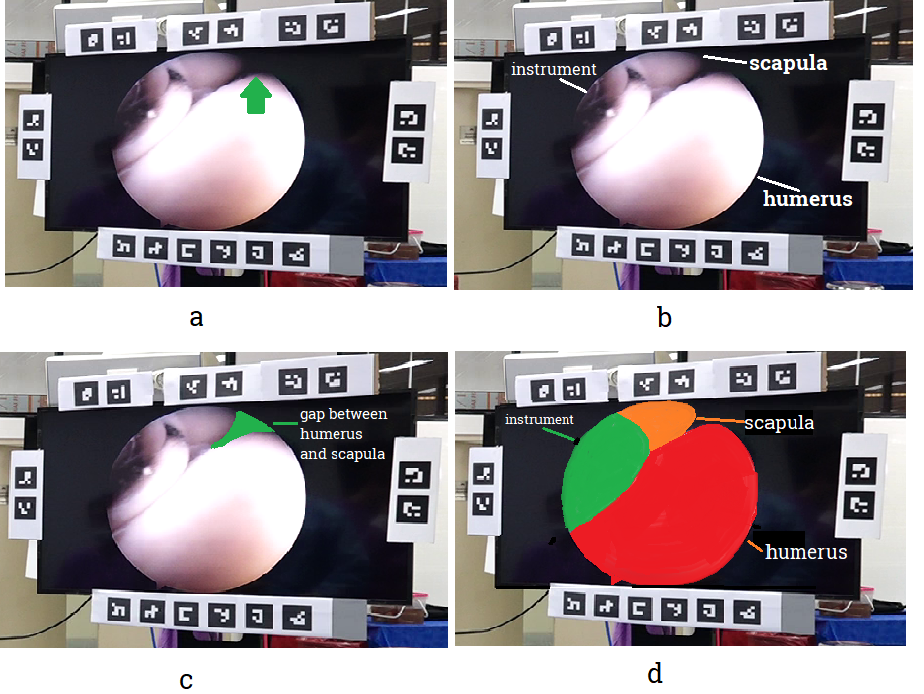}
		\caption{Different kinds of support for a confused surgeon.}
		
		\label{fig:support}
	\end{figure}

	The different kinds of support can be seen on Figure \ref{fig:support}. a) shows a simple arrow, which tells the surgeon where to go next with the arthroscope. b) shows more support by naming the single party of the output, so the surgeon knows which parts are involved and may remember how to proceed. Figure \ref{fig:support}, c shows a similar output like a), but there are only visual clues highlighted, and d) this help would provide the most support, by segmenting and coloring the single parts in different colors and name them, accordingly.

	\bibliographystyle{ACM-Reference-Format}
	\bibliography{acmart}
	
	\appendix
	
\end{document}